\documentclass[conference]{IEEEtran}
\IEEEoverridecommandlockouts
\usepackage{cite}
\usepackage{amsmath,amssymb,amsfonts}
\usepackage{algorithmic}
\usepackage{graphicx}
\usepackage{textcomp}
\usepackage{xcolor}

\usepackage{dsfont}
\usepackage{amsmath}
\usepackage{graphicx}
\usepackage{braket}
\usepackage[colorlinks=true, allcolors=blue]{hyperref}

\def\BibTeX{{\rm B\kern-.05em{\sc i\kern-.025em b}\kern-.08em
    T\kern-.1667em\lower.7ex\hbox{E}\kern-.125emX}}

\begin{document}

\title{C2QA - Bosonic Qiskit
\thanks{* These authors contributed equally.}
}

\author{
\IEEEauthorblockN{Timothy J Stavenger*}
\IEEEauthorblockA{\textit{Pacific Northwest National Laboratory}\\
Richland, WA, USA \\
timothy.stavenger@pnnl.gov \\
0000-0002-4270-5952}
\and
\IEEEauthorblockN{Eleanor Crane*}
\IEEEauthorblockA{
\textit{Joint Quantum Institute \& QuICS}\\\textit{NIST/University of Maryland}\\
College Park, MD, USA \\
ella@ellacrane.com \\
0000-0002-2752-6462}
\and
\IEEEauthorblockN{Kevin C Smith}
\IEEEauthorblockA{\textit{Brookhaven National Laboratory} \\
\textit{Yale University} \\
New Haven, CT, USA \\
kevin.smith@yale.edu \\
0000-0002-2397-1518}
\and
\IEEEauthorblockN{Christopher T Kang}
\IEEEauthorblockA{\textit{University of Washington} \\ \textit{Pacific Northwest National Laboratory} \\
Seattle, WA, USA \\
ck32@uw.edu \\
0000-0003-0105-7677}
\and
\IEEEauthorblockN{Steven M Girvin}
\IEEEauthorblockA{\textit{Yale University} \\
New Haven, CT, USA \\
steven.girvin@yale.edu \\
0000-0002-6470-5494}
\and
\IEEEauthorblockN{Nathan Wiebe}
\IEEEauthorblockA{\textit{University of Toronto} \\
\textit{Pacific Northwest National Laboratory} \\
Toronto, Canada \\
nathanwiebe@gmail.com \\
0000-0001-7642-1061}}
\maketitle

\begin{abstract}
The practical benefits of hybrid quantum information processing hardware that contains continuous-variable objects (bosonic modes such as mechanical or electromagnetic oscillators) in addition to traditional (discrete-variable) qubits  have recently been demonstrated by experiments with bosonic codes that reach the break-even point for quantum error correction \cite{ofek_extending_2016,hu_quantum_2019,ma_error-transparent_2020,Campagne2020,IonTrapGKP_QEC} and by efficient Gaussian boson sampling simulation of the Franck-Condon spectra of triatomic molecules \cite{wang2020efficient} that is well beyond the capabilities of current qubit-only hardware. The goal of this Co-design Center for Quantum Advantage (C2QA) project is to develop an instruction set architecture (ISA) for hybrid qubit/bosonic mode systems that contains an inventory of the fundamental operations and measurements that are possible in such hardware. The corresponding abstract machine model (AMM) would also contain a description of the appropriate error models associated with the gates, measurements and time evolution of the hardware. This information has been implemented as an extension of  Qiskit. Qiskit is an open-source software development toolkit (SDK) for simulating the quantum state of a quantum circuit on a system with Python 3.7+ and for running the same circuits on prototype hardware within the IBM Quantum Lab. We introduce the Bosonic Qiskit software to enable the simulation of hybrid qubit/bosonic systems using the existing Qiskit software development kit \cite{bosonic_qiskit_git}. This implementation can be used for simulating new hybrid systems, verifying proposed physical systems, and modeling systems larger than can currently be constructed. We also cover tutorials and example use cases included within the software to study Jaynes-Cummings models, bosonic Hubbard models, plotting Wigner functions and animations, and calculating maximum likelihood estimations using Wigner functions.
\end{abstract}

\begin{IEEEkeywords}
quantum, boson, Qiskit
\end{IEEEkeywords}


\section{Introduction}
When trying to solve physical questions of a quantum nature there is an obvious benefit in using quantum model systems to find the solutions. Even questions which are not inherently quantum mechanical, such as those involving the diagonalization of large matrices, is proposed to be much faster when mapped and solved using quantum systems~\cite{HHL}. In recent years, significant advances have been made towards developing tools which make use of the simplest quantum objects: two-level systems, or qubits. However, many problems are bosonic in nature: they require infinite-level systems\footnote{Not to be confused with superposition. A qubit can be in a superposition between its two levels, and a bosonic mode can be in a superposition between some or all of its levels}, for example relevant high-energy field theories, theories modelling photons or phonons, and certain topological models.

Although it is possible to approximate infinite-level systems with a tensor product of two-level ones by truncating the infinite-level system into a reduced multi-level system (choosing a cutoff), this is highly inefficient and the complexity of operations scales badly with cutoff as is shown in Girvin et al.~\cite{Girvin_ISA}. It is more natural and hardware-efficient to directly use multi- or infinite-level hardware. Due to rapid recent progress in the field of quantum circuit error detection (QED), using the many levels of microwave oscillator modes for continuous variable quantum computation is well on the way to becoming a reality~\cite{blais2021circuit}.

A major challenge facing the development of algorithms and applications for models of computing that hybridize ordinary qubits with bosonic modes is that it is difficult to express programs in existing languages.  This in part is because of the fact that the standard logical abstractions of boolean logic do not easily apply here.  New programming concepts are therefore needed to spur the development and compilation of quantum algorithms in this setting.

\begin{figure}
    \centering
    \includegraphics[width=0.4\textwidth]{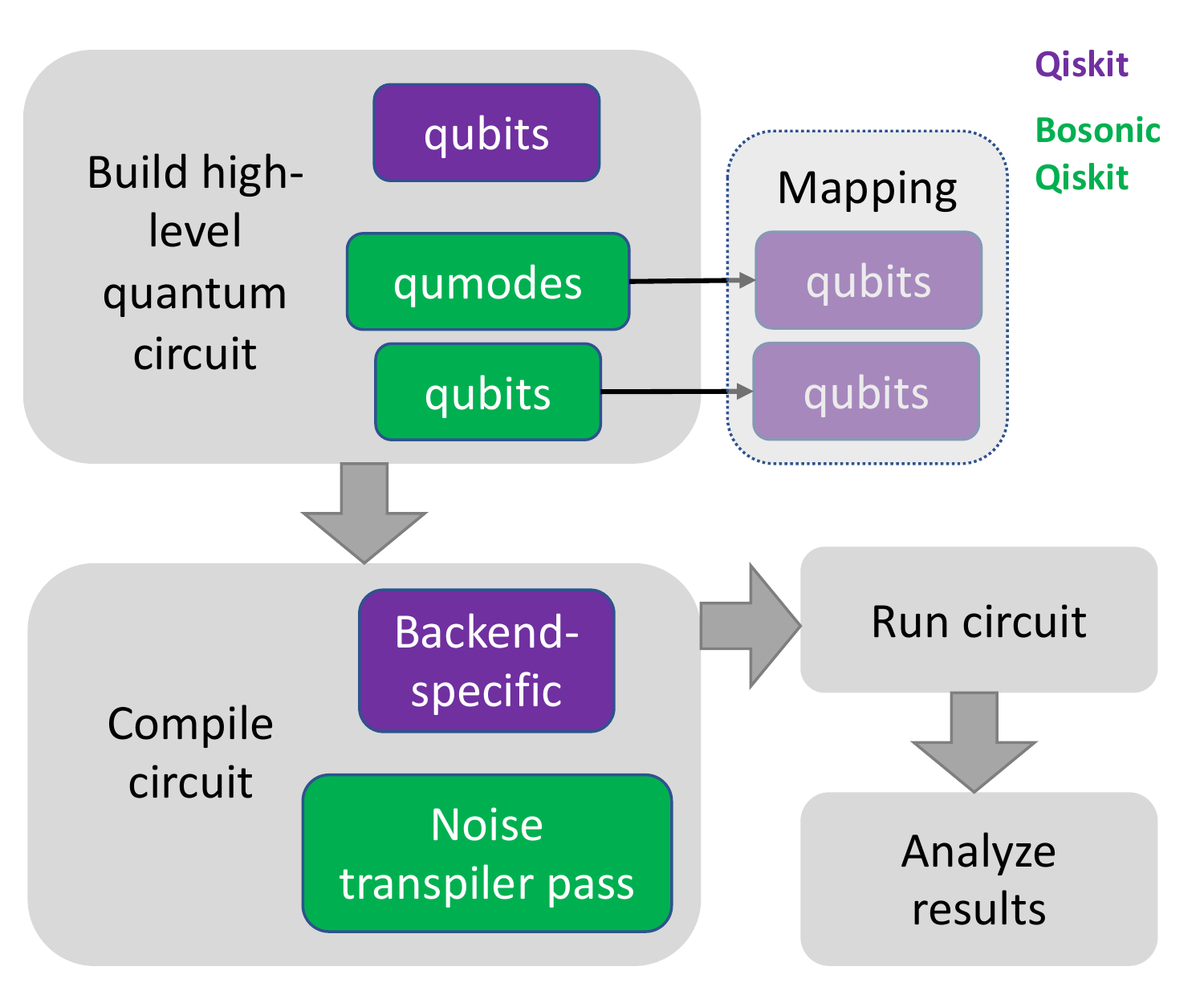}
    \caption{Block diagram of algorithm to transform and execute a quantum algorithm, comparing stock Qiskit to Bosonic Qiskit.}
    \label{fig:block_diagrams}
\end{figure}

In this work, we provide software simulation support for this hardware, building on an existing extremely successful open-source software development kit for qubit hardware,  Qiskit~\cite{noauthor_qiskit_nodate}, as represented in Fig.~\ref{fig:block_diagrams}. The Qiskit simulator extension is used as a way to simulate the bosons, not as a way to run the bosonic circuits on qubit hardware. This allows researchers to simulate potential bosonic circuits without needing to run actual qumode systems.

Qiskit is an open-source software development kit for simulating and executing quantum circuits using qubits. The Qiskit software can be used for both simulating circuits on classical systems as well as executing circuits on either IBM hosted superconducting quantum computers \cite{noauthor_qiskit.org_nodate} or any other hardware supporting \verb|QasmQobj|~\cite{cross_open_2017,noauthor_qasmqobj_nodate}. Qiskit uses the Python programming language to construct quantum circuits, compile them to a specified architecture, and either simulate or execute them on hardware. The use of Python gives researchers a wealth of integration opportunities with many other Python packages. Details for use of the Qiskit SDK can be found in the online documentation \cite{noauthor_qiskit_nodate}.

\section{Hybrid Qubit/Continuous-Variable Quantum Model}
The computational model that we consider involves both qubits and bosonic modes. In the rest of the paper we will refer to bosonic modes as modes or qumodes in the context of Bosonic Qiskit.  While qubit operations may be familiar to the reader, we review here the properties of the bosonic operations and common operations on the objects.  Our Hilbert space in this context is formally a tensor product of the form $\mathcal{H} = \mathcal{H}_{\rm qubit}^{m}\otimes \mathcal{H}_{\rm qumode}^{n}$ where $m$ and $n$ are the number of qubits and bosonic modes respectively.  We assume that a complete set of standard gate operations are provided on the qubits such as Hadamard, $R_z$ and CNOT.

To understand bosonic operations, it is helpful to introduce some notation. The occupation of the bosonic mode is an integer value referred to as the boson (or excitation) number. A state with definite boson number is known as a Fock state. For example, the Fock state with no occupancy is known as the vacuum state, denoted $\ket{0}$. Similarly, the state with $n$ bosons is denoted $\ket{n}$. Transitions between different Fock states are facilitated by $a^\dagger$, known as the creation (or raising) operator, and its Hermitian adjoint $a$, the annihilation (or lowering) operator. When acting on a Fock state, the creation operator increments the boson number by one, $a^\dagger\ket{n} = \sqrt{n+1}\ket{n+1}$, and its adjoint lowers the boson number, $a\ket{n}=\sqrt{n}\ket{n-1}$. The number operator\footnote{To avoid confusion, we will use a hat only to distinguish the number operator from the scalar $n$ throughout this text. For all other operators, no hat will be used.} $\hat{n}:=a^\dagger a$ returns the occupancy of the mode, $\hat{n} \ket{n} = n \ket{n}$.  We assume here the existence of a measurement operation  \cite{wang2020efficient} that is boson number resolving, meaning that on measurement it will return a specific occupation level of the measured mode.  

Because the number of Fock states is countably infinite, it is also possible to represent the state of a bosonic mode using a wave function that is continuous in the position $x$, or in the momentum $p$.  The Wigner function (described below) is a quasi-probability distribution in the oscillator phase space $W(x,p)$ that contains the same information as the density matrix $\rho(x,x^\prime)$. Such a phase space description of a qumode is also a useful visualization tool -- we refer the reader to the illustration in Fig~\ref{fig:cutoff}, where we plot the Wigner quasiprobability distribution for both a Fock state and a displaced vacuum state. In the former case, the number of nodal rings encircling the origin ($x=0$, $p=0$) corresponds to the boson occupation number. For the latter, it is possible to displace the qumode in any phase space direction using the displacement operator: $e^{\theta a^{\dagger} - \theta^* a}$. The resulting displaced state is called a \textit{coherent state}.

A natural platform for hybrid bosonic/qubit computations is circuit QED, in which the bosonic modes of microwave resonators are coupled to superconducting qubits. Other architectures involving both bosonic and qubit degrees of freedom have also been developed, e.g. using phononic modes in trapped ion systems \cite{IonTrapGKP_QEC,C7SC04602B} and optical modes in photonic platforms \cite{Huh_BosonSampling_Optical}. Bosonic Qiskit is not limited to any specific hardware, though we emphasize that its current implementation primarily includes built-in gates which have been demonstrated in the circuit QED platform.

Standard gate operations on the bosonic modes are given in Table~\ref{tab:gates}.  The top section of the table lists Gaussian gates alone, which - combined with photon resolving measurement - are universal for quantum computing \cite{Girvin_ISA}. The bottom section includes a selection of hybrid qubit-bosonic non-Gaussian operations.

As an example of the latter, consider the SNAP (Selective Number-dependent Arbitrary Phase) gate which, conditioned on the state of the ancilla qubit, applies a different programmable phase, $\theta_n$, to each Fock state: ${\rm SNAP}(\theta) \ket{\psi}\ket{n} \mapsto e^{-i\sigma^z\theta_n}\ket{\psi}\ket{n}$, where $\ket{\psi}$ is the state of the qubit. It can be used to operationalize measurements on the bosonic mode and can also be paired with single qubit operations and Gaussian bosonic operations to achieve universal control without the use of measurement and feed forward \cite{heeres2015cavity}.

A practical concern is that, in practice, we often need to truncate the Hilbert space to perform a simulation. The aim of our software is to implement generic bosonic operations and tight integration with Qiskit while simplifying the management of issues involving cutoffs.

\section{Bosonic Qiskit}

The Bosonic Qiskit software package represents the first $2^k$ levels of a bosonic mode (qumodes) as a register of $k$ qubits within the Qiskit software with a binary encoding representing the Fock state\footnote{Various other representations of continuous variables with two-level systems also exist such as~\cite{Jordan_Lee_Preskill}.}, which can be used in conjunction with qubits and classical bits. The resulting representation is then simulated on a classical system in an analogous fashion to base Qiskit circuit\footnote{Note that although Bosonic Qiskit circuits can be simulated in software, they cannot be directly run on qubit hardware without further compilation given that the gates are designed for hybrid qubit-qumode hardware.}. 

A qubit is a two-level system, hosting a spin-up and spin-down state, whereas a qumode can theoretically host an infinite number of bosons. A qumode state of definite boson number or occupation is called a Fock state. The qumodes are represented in a register using the \verb|QumodeRegister| class, which is a wrapper of Qiskit's \verb|QuantumRegister| class. This \verb|QumodeRegister|, along with any \verb|QuantumRegister| and \verb|ClassicalRegister|, is used to instantiate the custom circuit class, \verb|CVCircuit|, which extends Qiskit's \verb|QuantumCircuit|. The \verb|CVCircuit| class is where all custom bosonic gates are implemented. A benefit of extending existing classes is that those already familiar with programming in Qiskit should find Bosonic Qiskit familiar as well. In addition, any code which would work with the base Qiskit software will also work with the Bosonic Qiskit software package.

The following section gives a brief overview of various functionalities implemented into the Bosonsic Qiskit software and some notes to consider when using it. We take the probabilistic preparation of a cat state in one qumode using one ancilla qubit as a guiding example throughout. Section \ref{tutorials} gives an overview of a select number of Python Notebook tutorials present in \url{https://github.com/C2QA/bosonic-qiskit/tree/main/tutorials}.

\subsection{Summary of the guiding example}

In the following, we will initialize a circuit, add gates, and measure and visualize the state of a qubit and qumode, respectively, all within the context of a simple example: the preparation of a cat state. Importantly, this preparation is non-deterministic in the sense that the parity of the cat will be probabilistically determined upon measurement of an ancilla qubit. 

While the complete implementation will be demonstrated in the following, here we briefly summarize the core idea. The direction of the displacement of a qumode can be conditioned on the state of an ancilla qubit via the controlled displacement operator: $e^{\sigma^z\otimes \theta a^{\dagger} - \theta^* a}$. If the qubit is initially placed in an equal superposition of $\sigma^z$ eigenstates (using e.g. the Hadamard gate), the mode will be displaced in a different direction depending on the state of the qubit, thereby entangling the qumode with the qubit. If the qubit is then measured in the $\sigma^x$ basis, the qumode will, depending on the measurement outcome, collapse onto either an odd- or even-parity superposition of coherent states known as a cat state. The circuit diagrams and corresponding Wigner functions are shown in Fig.~\ref{fig_cat}.

\begin{figure}
    \centering
    \includegraphics[width=0.5\textwidth]{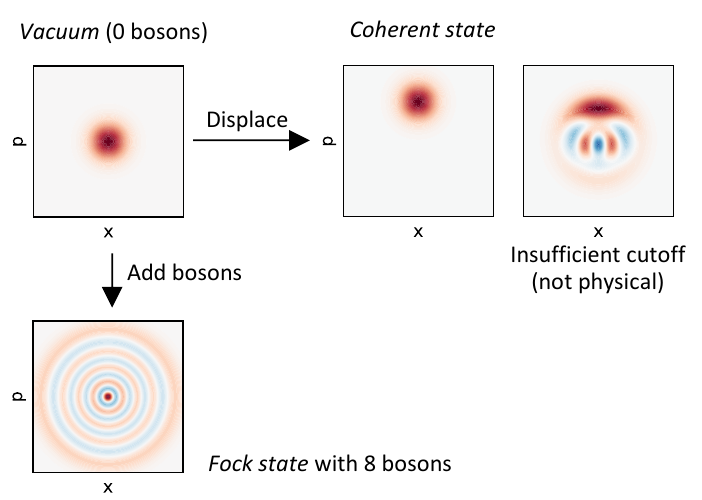}
    \caption{\textbf{Bosonic Qiskit illustration of the basic Wigner functions} (state of the qumode). $p$: momentum, $x$: position. Top left: vacuum state. Bottom: Fock state 8. Middle: coherent state. Top right: un-physical state. This illustrates that displacing the vacuum in momentum using a cutoff of 6 qubits per qumode (expected result, left) and a cutoff of 2 qubits per qumode (artifacts, right). Red corresponds to positive and blue negative values.}
    \label{fig:cutoff}
\end{figure}

\subsection{Instantiating a bosonic circuit}

To create a Bosonic Qiskit circuit, we can concatenate a \verb|QumodeRegister|, \verb|QuantumRegister|, and \verb|ClassicalRegister| into a \verb|CVCircuit| class. The example code below instantiates the qumode in which the cat state will be created, the qubit used to help create it, and a classical register to read out measurement results:

\begin{verbatim}
    qmr = c2qa.QumodeRegister(
        num_qumodes=1, 
        num_qubits_per_qumode=6)
    qbr = qiskit.QuantumRegister(1)
    cr = qiskit.ClassicalRegister(1)
    circuit = c2qa.CVCircuit(qmr, qbr, cr)
\end{verbatim}
    
Note that even though qumodes are represented by collections of qubits within Qiskit, the qumodes within the \verb|QumodeRegister| are still addressed as individual qumodes, analogous to individual qubits in a \verb|QuantumRegister|. The Bosonic Qiskit software abstracts away operational details at the level of the underlying qubits, allowing users to simulate bosonic operations on qumodes directly without having to consider the composing qubits.

\subsection{Adding gates to the circuit}

Following the implementation details found in ``Instruction Set Architecture and Abstract Machine Models for Hybrid Qubit/Continuous-Variable Quantum Processors''~\cite{Girvin_ISA}, the gates in Tab.~\ref{tab:gates} have been implemented in Bosonic Qiskit. In these examples, $a^{\dagger}$ ($a$) and $b^{\dagger}$ ($b$) refer to creation (annihilation) operators on two different modes, and ${\sigma^z}$ is the Pauli Z single qubit operation. For implementation details, see the \verb|c2qa/operators.py| and \verb|c2qa/circuit.py| modules within Bosonic Qiskit.

\begin{table}
\caption{Bosonic gates implemented in Bosonic Qiskit.\\ Top: Gaussian gates. Bottom: non-Gaussian gates. }
\begin{center}
\begin{tabular}{ |c|c|c| } 
 \hline
Phase space rotation & $e^{i\theta \hat{n}}$ & \verb|cv_r()|\\
Displacement & $e^{\theta a^{\dagger} - \theta^* a}$ & \verb|cv_d()|\\ 
Single-mode squeezing & $e^{\frac{1}{2}(\theta^* a a - \theta a^{\dagger}a^{\dagger})}$ & \verb|cv_sq()|\\ 
Two-mode squeezing & $e^{(\theta^* a b - \theta a^\dagger b^\dagger)}$ & \verb|cv_sq2()|\\
Beamsplitter & $e^{\theta a^{\dagger}b-\theta^* b^{\dagger}a}$ & \verb|cv_bs()|\\
\hline
Controlled rotation & $e^{\sigma^z \otimes i \theta \hat{n}}$ & \verb|cv_c_r()|\\
Controlled displacement & $e^{\sigma^z\otimes (\theta a^{\dagger} - \theta^* a)}$& \verb|cv_c_d()|\\
Controlled beam-splitter & $e^{\sigma^z\otimes ( \theta a^{\dagger}b-\theta^*b^{\dagger}a)}$ & \verb|cv_c_bs()|\\
(Controlled) SNAP & $e^{\sigma^z  \otimes i \theta_n \ket{n}\bra{n}}$ & \verb|cv_snap()|\\
Exponential SWAP & $e^{i \frac{\theta}{2} \mathrm{SWAP}}$ & \verb|cv_eswap()|\\
\hline
\end{tabular}
\end{center}
\label{tab:gates}
\end{table}

\textbf{Gaussian and non-Gaussian gates.} Much like Clifford operations must be combined with non-Clifford gates to achieve universal operations on a single qubit, the strength of bosonic circuits lies in the capability to combine Gaussian operations with non-Gaussian, qubit controlled qumode gates. In Table \ref{tab:gates}, we divide the implemented gates according to these two categories, with the top and bottom rows enumerating Gaussian and non-Gaussian operations, respectively.

\textbf{Qumode gates.} The phase space rotation gate rotates the qumode in phase space by a specified angle. The displacement gate displaces the qumode in an amount and direction specified by its (complex) parameter. The single-mode squeezing gate creates and destroys photons in such a way that it diminishes fluctuations along one phase space quadrature at the expense of increasing fluctuations along the orthogonal quadrature. Similarly, the two-mode squeezing gate creates and destroys pairs of photons, one in each qumode, such that their fluctuations become correlated. The beamsplitter gate facilitates exchange of quanta between cavities.

\textbf{Qubit controlled and conditional qumode gates.} Many of the gates listed above can be controlled, e.g. the controlled displacement (\verb|cv_c_d(|$\theta$\verb|,qma,qb)|, where $\theta$ is a variable parameter, \verb|qma| is a qumode, and \verb|qb| is a qubit). A particularly powerful example is the previously introduced SNAP gate, which applies a chosen phase to a particular Fock state. We also include the exponential SWAP gate, a useful entangling operation \cite{gao_entanglement_2019} which applies a weighted superposition of identity and SWAP gates to two qumodes. Many other useful gates can be synthesized through combination of these listed. One example is the controlled parity operation $e^{i \frac{\pi}{2} [(\sigma^z + \mathds{1})\otimes \hat{n}]}$, created by combining a phase space rotation with a controlled rotation. This gate rotates the qumode only if the state of the qubit is such that $\sigma^z\ket{\psi} = \ket{\psi}$.


\textbf{Adding the gates.} To add gates to the circuit, gates are appended to the \verb|CVCircuit| using the helper functions implemented in the \verb|c2qa/circuit.py| module - similar to native Qiskit. We do this below in the context of the non-deterministic cat creation, following the code snippet in the previous section. We can first initialize the qumodes in the \verb|QumodeRegister| to a certain Fock state. We choose the initial state of the first and only qumode in the register to be the vacuum (so one does not technically need to initialize the qumode):
\begin{verbatim}
    circuit.cv_initialize(0, qmr[0])
\end{verbatim}
The first input is an integer denoting the Fock state $\ket{n}$ to initialize (in this case, $\ket{0}$). While not needed for the present demonstration, \verb|circuit.cv_intialize| can also prepare a superposition of Fock states; this is achieved by passing a \verb|List| whose $i$th entry is the complex amplitude of Fock state $\ket{i}$. 

\begin{figure}
    \centering
    \includegraphics[width=0.5\textwidth]{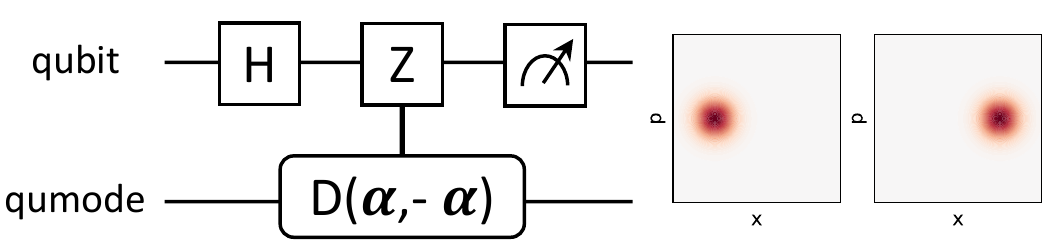}\\
    \includegraphics[width=0.5\textwidth]{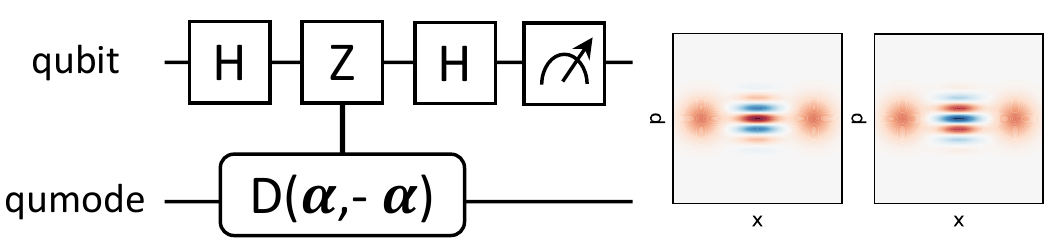}
    \caption{Bosonic Qiskit circuit and Wigner functions showing the stages of preparation of a non-deterministic cat state. In the upper panel, the qumode is projected onto either the $\ket{+\alpha}$ or $\ket{-\alpha}$ coherent state, depending on the outcome of the qubit measurement ($\ket{0}$ or $\ket{1}$) following the controlled displacement. In the lower panel, the qubit is first transformed using a Hadamard gate prior to measurement, and the qumode is thus projected onto an even ($\ket{0}$) or odd ($\ket{1}$) cat state upon measurement of the ancilla. These two outcomes can be differentiated by the color of the fringes.}
    \label{fig_cat}
\end{figure}

Next, we put the qubit into a superposition using the Hadamard gate, and then displace the vacuum controlled on the state of the qubit, as is shown in Fig.~\ref{fig_cat}:
\begin{verbatim}
    alpha = 1
    circuit.h(qbr[0])
    circuit.cv_c_d(
        alpha, qmr[0], qbr[0])
\end{verbatim}
The index \verb|0| correspond to the specific qubit or qumode on which the gate should be performed (in this example there is only one qubit in the \verb|QubitRegister| and only one qumode in the \verb|QumodeRegister|). Finally, we measure the qubit in the $\sigma^x$ eigenbasis, 
\begin{verbatim}
    circuit.h(qbr[0])
    circuit.measure(qbr[0],cr[0])
\end{verbatim}
the result of which will determine the parity of the cat state, as shown in Fig. \ref{fig_cat},


\subsection{Simulating the circuit}
It is possible to simulate the circuit using the Qiskit Aer simulator with the \verb|simulate(circuit)| function inside the \verb|c2qa/util.py| module. This function handles adding noise model transpiler passes and returns the simulated state vector, if requested. Other simulators, like the the IBM Matrix Product State simulator freely accessible online, can be used as well.
Here is an example in which the state of a \verb|CVCircuit| is simulated:
\begin{verbatim}
    state, result =
        c2qa.util.simulate(circuit)
\end{verbatim}

\subsection{Qumode state readout and measurement}

\subsubsection{Simulator statevector}

With the Qiskit Aer simulator, the statevector (stateop) can be printed out after the simulation of a quantum circuit. The function within the \verb|c2qa/util.py| module: \begin{verbatim}
    cv_stateread(
        state, 
        qregister_list, 
        cregister_list)
\end{verbatim}
enables the printing to standard out, for each many-body configuration of a superposition, the Fock state of each qumode, the computational basis state of each qubit, and its complex amplitude.

\subsubsection{Measurement}

We also provide functionality to add measurement of both qubits and qumodes to a circuit. In Qiskit, qubit measurements may be appended to a circuit with the \verb|QuantumCircuit.measure| method. In \verb|circuit.py|, we supply a generalization of this method,
\begin{verbatim}
    cv_measure(
        self, 
        qubit_qumode_list, 
        cbit_list),
\end{verbatim}
which accepts a list of qubits and qumodes (\verb|qubit_qumode_list|) to be measured and mapped onto a list of classical bits (\verb|cbit_list|). Similar to base Qiskit, qubits are measured in the computational basis, while qumodes are measured in the Fock basis and mapped onto classical bits using a binary encoding. Measurement counts are reported in little-endian ordering with respect to \verb|qubit_qumode_list|. Separately, we provide the method \verb|cv_fockcounts(counts, qubit_qumode_list)| in \verb|util.py| which accepts a \verb|dict| of measurement counts (as returned by \verb|Result.get_counts()| -- see \url{https://qiskit.org/documentation/stubs/qiskit.result.Result.get_counts.html}) and returns a new \verb|dict| with Fock numbers represented as base-10 integers.

\subsection{Support for Wigner Functions}
\subsubsection{Plotting Wigner functions}

The Wigner function is a phase space quasiprobability distribution for a bosonic mode containing the same state tomography information as the density matrix.  The Python module in \verb|c2qa/util.py| includes a \verb|wigner()| method to calculate the Wigner function from a given Qiskit state vector simulation result. A qumode Wigner function can then be plotted either with or without projection of an ancilla qubit onto the basis states $\ket{0}$, $\ket{1}$, $\ket{+}$, and $\ket{-}$. The tutorial Python Notebook found in \verb|tutorials/wigner-function| steps a user through the use of both the method to calculate the Wigner function as well as the methods to plot the results. The result is a Matplotlib generated image of the provided state vector. An example which uses a circuit that simply creates a coherent state in the qumode can be seen in Fig. ~\ref{fig:cutoff}, while Fig.~\ref{fig_cat} shows the result of the nondeterministic cat preparation circuit demonstrated described above.



\subsubsection{Animating Wigner functions}
The Python module in \verb|c2qa/util.py| also includes an \verb|animate_wigner()| method to animate a circuit as a series of Wigner function plots saved frame-by-frame into either an animated GIF or an MP4 video. In both cases, evolution under the circuit is animated by incrementally applying the gates and saving the resulting Wigner function plot as a frame in the movie. The user has the ability to configure how many frames will be generated per gate, changing the number of frames and length of the resulting animation. As with plotting static Wigner functions, the \verb|tutorials/displacement-calibration| folder contains a calibration circuit. The result is an animated GIF or MP4, as specified by the user, of the calibration process.

\subsubsection{Reduced density matrices}
It is possible to obtain the state of only the qumodes or of only the qubits by using the reduced density matrix functionality. This corresponds to tracing out the states of the qumodes (qubits) which results in the state of only the qubits (qumodes). Using the \verb|c2qa/util.py| module, this can be called with \verb|cv_partial_trace()|.

\subsubsection{Maximum likelihood estimation}
As another method to calculate the Wigner function from a given state vector, the \verb|c2qa/util.py| module includes the \verb|wigner_mle()| function to calculate the maximum likelihood estimation (MLE) from an array of simulation state vectors (i.e., many Qiskit simulation shots). The implementation makes use of SciPy's statistical functions for normal  continuous random variables to perform the MLE calculation\cite{noauthor_scipy_nodate}. Just as before, the MLE results are then passed to the Wigner function method described above for calculation. Also as before, the Wigner function can be plotted using the \verb|c2qa/util.py| module's \verb|plot()| function. A tutorial exercising the MLE feature can be found in \verb|tutorials/wigner-mle|.

\subsection{Warnings}

\subsubsection{Choosing the qumode cutoff}

The cutoff is set to $2^{n}$, where $n$ is the number of qubits per mode (specified by the user when instantiating the \verb|QumodeRegister|). For 2 qubits per mode for example, we have for the boson annhilation and creation operators:
\begin{equation}
a=\begin{pmatrix}
0&1&0&0\\
0&0&\sqrt{2}&0\\
0&0&0&\sqrt{3}\\
0&0&0&0
\end{pmatrix}, \quad a^{\dagger}=\begin{pmatrix}
0&0&0&0\\
1&0&0&0\\
0&\sqrt{2}&0&0\\
0&0&\sqrt{3}&0\\
\end{pmatrix},
\end{equation}

Note that $a a^{\dagger}$ and $(a^{\dagger} a + 1 )$, which according to the commutation relation ($[a,a^{\dagger}]=1$) should yield the same result, do not:
\begin{align}
    a a^{\dagger}\ket{3} &= 0 \\
    (a^{\dagger} a + 1 ) \ket{3} &= 4.
\end{align}
The second expression is the expected result, indicating  that it is advantageous to normal order the operators to reduce the risk of running into the cutoff. The effects of cutoff are demonstrated in the Wigner Function notebook described below and reproduced in Fig.~\ref{fig:cutoff}. Displacements are particularly sensitive to the cutoff, but boson-number-preserving operations are not.

\subsubsection{Endianness}
In Qiskit, qubits are indexed and represented right-to-left, in little-endian order. This is especially important to note when creating custom operator matrices for new gates as well as interpreting measurement results and state vectors that are output from Qiskit simulations. The little-endian representation may be the opposite of what a user is expecting, given the way the qubits and qumodes are sent to the quantum circuit. To remedy this, several Bosonic Qiskit functions allow the user to choose between little- and big-endian ordering.

By default \verb|cv_fockcounts| will preserve the endianness of the counts \verb|dict| which is passed in, and therefore will be in little-endian ordering if retrieved using Qiskit's built-in \verb|Result.get_counts()| method. However, \verb|cv_fockcounts| accepts a flag parameter \verb|reverse_endianness| which, if \verb|True|, will convert from little-endian to big-endian (or vice-versa).
 The default behavior of \verb|cv_stateread| is to print basis states using little-endian ordering. Big-endian ordering may be specified by setting the parameter \verb|big_endian=True|. See discussion of these functions above.

\section{Use Cases and Tutorials}\label{tutorials}
We have provided a number of use cases and tutorials which can be found at \url{https://github.com/C2QA/bosonic-qiskit/tree/main/tutorials}. Two of them are summarized below.

So far, in this paper, much attention has been given to the manipulation and visualisation of a single qumode. However, Bosonic Qiskit is also well adapted to quantum simulation of dynamics of many-body systems. In the following, we demonstrate the capability to calculate the time evolution generated by two paradigmatic Hamiltonians involving bosonic modes, using Bosonic Qiskit.



\subsection{Jaynes-Cummings model in the dispersive regime}

The Jaynes-Cummings model describes a simple system of a qumode coupled to a two-level system. It is written below in the dispersive coupling parameter regime, a particularly relevant scenario for the circuit QED platform \cite{blais2021circuit}:
\begin{equation}
    H=\omega_\mathrm{R}a^\dagger a + \frac{\omega_\mathrm{Q}}{2}\sigma^z+\frac{\chi}{2}\sigma^z a^\dagger a\label{eq:dispersiveHD2}.
\end{equation}
To simulate this Hamiltonian, a prerequisite is to implement the time evolution operator\footnote{For simplicity, $\hbar$ is taken to be 1.} $\mathcal{U}=e^{-i H t}$. Because all terms commute with one another, the task of realizing $\mathcal{U}$ is reduced to the implementation of the exponential of each Hamiltonian term independently. This can be achieved using the previously defined gates as summarized below:
\begin{equation}
    \begin{split}
        \omega_R a^\dagger a &\rightarrow e^{-i\omega_R t a^\dagger a} \quad\textrm{(phase space rotation)} \\
        \omega_Q \sigma^z/2 &\rightarrow e^{-i\omega_Q t \sigma^z/2} \quad\textrm{(Qiskit $R_z$ gate)} \\
        \chi a^\dagger a &\rightarrow e^{-i\chi t \sigma^z a^\dagger a/2} \quad\textrm{(controlled phase space rotation)}
    \end{split}
\end{equation}

\subsection{Bose-Hubbard model}

The Bose-Hubbard model describes spinless bosons hopping on a lattice. Its most interesting feature is the superfluid-to-Mott-insulator transition~\cite{greiner_quantum_2002}. The Hamiltonian is written as follows:
\begin{equation}
H=-J\sum_{\braket{ij}}\left( a_i^{\dagger}a_j + \mathrm{h.c.} \right) + \frac{U}{2} \sum_i \hat{n}_i \left( \hat{n}_i - 1 \right) - \mu \sum_i \hat{n}_i
\end{equation}
where $\braket{ij}$ describes summation over neighbouring lattice sites, $J$ denotes the strength of the hopping, $U$ is the on-site interaction, $\mu$ is the chemical potential, and $\hat{n}_i = a_i^\dagger a_i$ is the number operator for the $i$th site. 

\begin{figure}
    \centering
    \includegraphics[width=0.5\textwidth]{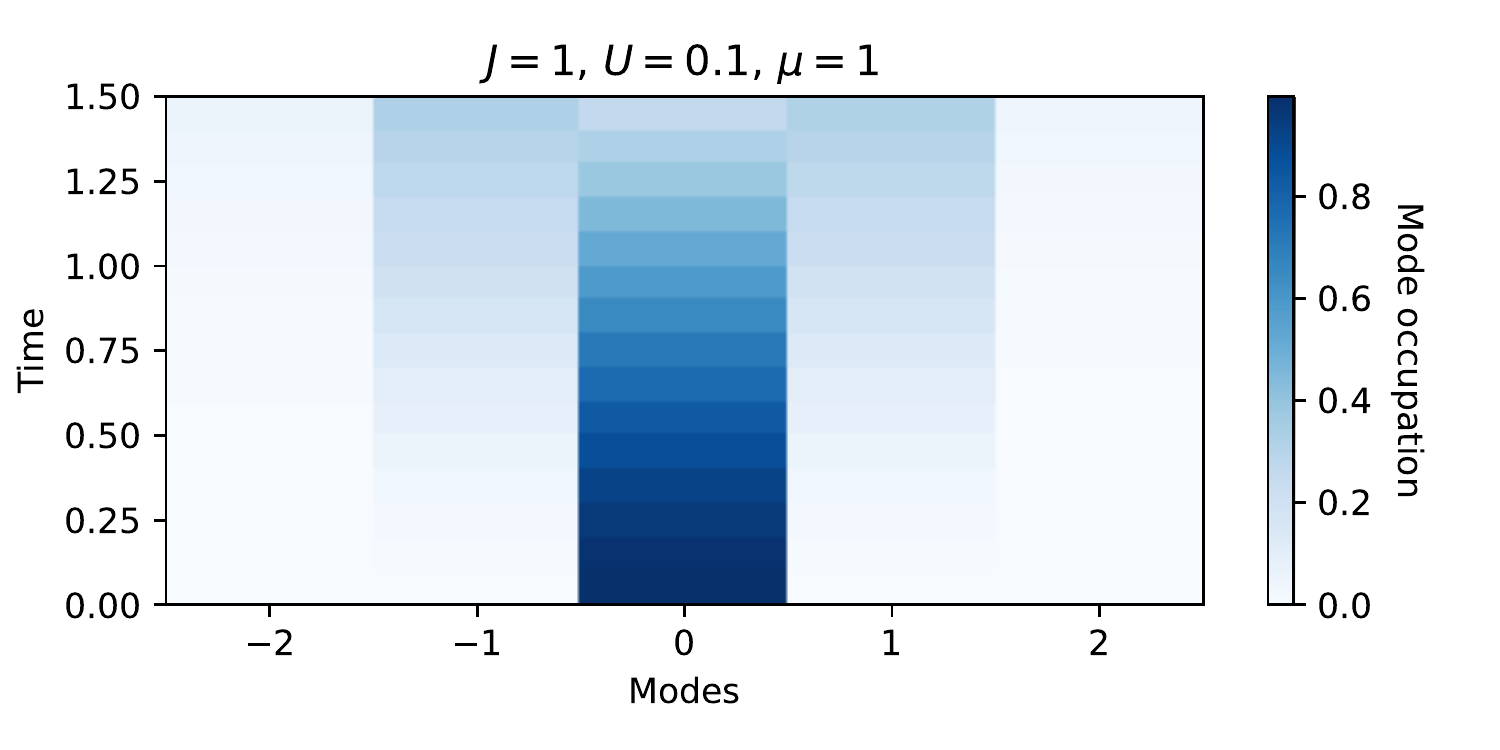}
    \caption{Dynamics of the Bose-Hubbard model in the regime of large hopping and small on-site and chemical potential terms, generated with Bosonic Qiskit.}
    \label{fig:BH}
\end{figure}

Using Bosonic Qiskit, we study the Trotterized time evolution of the occupations of the Bose-Hubbard model with five sites, starting from an initial state containing one boson in the central mode. Similar to the Jaynes-Cummings tutorial, the goal is to implement the time evolution operator $\mathcal{U}=e^{-i H_j dt}$ for each term of the Hamiltonian $H_j$, where $dt$ corresponds to the time evolution step. 

Among the three sets of terms, two are straightforward: each hopping term can be implemented using a beamsplitter between neighboring sites with parameter $\theta = - i J dt$ and the chemical potential term contributes a constant to the energy that simply generates a phase space rotation. Time evolution under the on-site interaction terms proportional to $\hat{n}_i(\hat{n}_i-1)$ can be directly implemented using SNAP gates. Alternatively, it is possible to synthesize a suitable gate using the Baker-Campbell-Hausdorff formula and (appropriately rotated) phase-space conditional rotation gates:
\begin{align}
e^{i\theta\sigma^x\hat{n}_j}e^{i\theta\sigma^y\hat{n}_j}e^{-i\theta\sigma^x\hat{n}_j}e^{-i\theta\sigma^y\hat{n}_j}\notag  =&e^{-\theta^2[\hat{n}_j\sigma^x,\hat{n}_j\sigma^y]+O(\theta^3)}\notag\\
=&e^{-i2\theta^2 \sigma^z \hat{n}_j\hat{n}_j+O(\theta^3)}.\label{eq_BCH}
\end{align}
Choosing $\theta = \sqrt{(U/4)\,dt}$ and noting that $e^{-i \phi \hat{n}_j(\hat{n}_j-1)}=e^{i\phi\hat{n}_j}e^{-i \phi\hat{n}_j^2}$, this strategy allows for simulation of the requisite term with Trotter errors that scale as $dt^{3/2}$, using a control qubit initialized to the state $\ket{0}$. 

\subsection{Use of IBMQ simulators}
It is possible to simulate a \verb|CVCircuit| using IBM Quantum (IBMQ) cloud-based service. For instructions on accessing IBMQ simulators, see details at \url{https://quantum-computing.ibm.com/lab/docs/iql/manage/account/ibmq}.

\section{GitHub}
\subsection{Continuous integration and delivery}
The Bosonic Qiskit repository uses GitHub Workflows to test the implemented functionality on each Git push. Test cases can be found in \url{https://github.com/C2QA/bosonic-qiskit/tree/main/tests} and use the PyTest software package \cite{noauthor_pytest:_nodate}. Within the GitHub workflow, these tests are run as a matrix build in virtual Linux, MacOS, and Windows systems across Python 3.7, 3.8, 3.9, and 3.10. 

On release of a new version in GitHub, another workflow automatically tests, packages, and publishes the new package to PyPI at \url{https://pypi.org/project/bosonic-qiskit/}. This allows other users to start using Bosonic Qiskit as a PyPI dependency rather than needing the source directly.

\subsection{Contributing to Bosonic Qiskit}\label{subsec:contributing}

With a general understanding of Bosonic Qiskit from above, it is possible to add custom gates representing bosonic operations. 

Note that the Bosonic Qiskit code is structured to separate generation of the operator matrices from creating instances of Qiskit Gate, in \verb|c2qa/operators.py| and \verb|c2qa/circuit.py| files, respectively. The first step in adding a new gate is to develop software to build a unitary operator matrix. These matrices must be unitary in order for Qiskit to simulate them. Non unitary matrices will fail during simulation. Existing operator matrices are built in the \verb|CVOperators| class found in \verb|c2qa/operators.py|. Included in \verb|CVOperators| are the user-specified cutoff, number of qumodes, as well as the bosonic creation and annihilation operators. The order of the data in your operators must match the order of the qumodes (Qiskit qubits) sent in as Qiskit gate parameters found in \verb|circuit.py|, as described next.

Once the software is written to build the operator matrix, a new function is added to the \verb|CVCircuit| class found in \verb|c2qa/circuit.py|. This class extends the Qiskit \verb|QuantumCircuit| class to add the bosonic gates available in this library. The previously defined operators are parameterized by user input, as needed, and appended to the \verb|QuantumCircuit| as unitary gates. The \verb|CVCircuit| class includes functions to easily make your new gates conditional based on a control qubit.

\subsubsection{Ensuring that operations are unitary}
Due to the use of \verb|UnitaryGate| in Bosonic-Qiskit, all custom bosonic operations performed must also be unitary. Creating new custom gates with a non-unitary operator will produce errors from Qiskit upon simulation. The circuit will fail on simulation until the operator is defined as a unitary matrix. Care must be taken when defining new custom operators to ensure proper values.

See the \verb|is_unitary_matrix()| function implementation in \url{https://github.com/Qiskit/qiskit-terra/blob/main/qiskit/quantum_info/operators/predicates.py} for details on Qiskit unitary matrix validation.

\section{Conclusions and Future Work}

We have introduced Bosonic Qiskit, an extension of the Qiskit software development kit. Bosonic Qiskit adds simulation support to Qiskit for continuous variable, bosonic quantum systems. The existing Bosonic Qiskit implementation includes many custom bosonic gates, a Wigner function visualization tool, and several tutorials and use cases.

Looking forward, this work is a first step towards the development of a comprehensive software stack for hybrid qubit systems such as those that appear in cavity QED.  Subsequent work, such as optimizing compilers at a gate or pulse level and developing Qiskit transpilers to run the bosonic simulations on qubit hardware, will be needed in order to realize the full potential of this emerging technology and further tools will be needed to abstract away error correction that will be used on top of such physical descriptions.  Such a stack capable of accepting a high level algorithmic description and yielding a sequence of operations appropriate for implementation will not only be a major step forward for the technology but also will take us closer to our ultimate goal of building a practical and scalable quantum computer.


\subsection{Bosonic Noise Modeling}
Efforts are currently underway to implement a custom Qiskit transpilation pass to add duration-dependent noise to circuits. With durations applied to each gate, this will allow bosonic noise (e.g., photon loss modeled as Kraus operators) to be continuously modeled throughout the simulated circuit execution. This will add considerable realism to the bosonic circuits not available in the existing implementation. To learn details of the current noise modeling, see the \verb|bosonic-noise-model| Git branch at \url{https://github.com/C2QA/bosonic-qiskit/tree/bosonic-noise-model}.

\subsection{Parameterized Circuit Simulation}
In an effort to better support parameterized circuits (see \url{https://qiskit.org/documentation/tutorials/circuits_advanced/01_advanced_circuits.html#Parameterized-circuits}), work is ongoing to add similar parameterized circuit support to Bosonic Qiskit. Among other benefits, this will ease integration of Bosonic Qiskit's gates into VQE circuits. For details on the current implementation status, see the \verb|parameterized-gates| branch at \url{https://github.com/C2QA/bosonic-qiskit/tree/parameterized-gates}.

\section*{Acknowledgements}
This project was supported by the U.S. Department of Energy, Office of Science, National Quantum Information Science Research Centers, Co-design Center for Quantum Advantage under contract number
DE-SC0012704.

Eleanor Crane was supported by UCL Faculty of Engineering Sciences and the Yale-UCL exchange scholarship from RIGE (Research, Innovation and Global Engagement), and by the Princeton MURI award SUB0000082, the DoE QSA, NSF QLCI (award No.OMA-2120757), DoE ASCR Accelerated Research in Quantum Computing program (award No.DE-SC0020312), NSF PFCQC program.

\bibliography{bib.bib}
\bibliographystyle{unsrt}

\end{document}